\documentclass[prb,twocolumn,showpacs,a4paper]{revtex4}
\usepackage{graphicx}% Include figure files

\begin{document}

\title{Strain distribution in quantum dot of arbitrary polyhedral
shape: \\ Analytical solution in closed form}

\author{A.V. Nenashev}
\email{nenashev@isp.nsc.ru}
\author{A.V. Dvurechenskii}
\affiliation{Institute of Semiconductor Physics, 630090,
Novosibirsk, Russia}
\affiliation{Novosibirsk State University,
630090, Novosibirsk, Russia}

\date{\today}

\begin{abstract}
An analytical expression of the strain distribution due to lattice
mismatch is obtained in an infinite isotropic
elastic medium (a matrix) with a three-dimensional polyhedron-shaped
inclusion (a quantum dot). The expression was obtained utilizing
the analogy between electrostatic and elastic theory problems. The
main idea lies in similarity of behavior of point charge electric
field and the strain field induced by point inclusion in the matrix.
This opens a way to simplify the structure of the expression for the
strain tensor. In the solution, the strain distribution consists of
contributions related to faces and edges of the inclusion. A
contribution of each face is proportional to the solid angle at
which the face is seen from the point where the strain is
calculated. A contribution of an edge is proportional to the
electrostatic potential which would be induced by this edge if it is
charged with a constant linear charge density. The solution is valid
for the case of inclusion having the same elastic constants as the
matrix. Our method can be applied also to the case of semi-infinite matrix with a free surface.
Three particular cases of the general solution are
considered---for inclusions of pyramidal, truncated pyramidal, and
``hut-cluster'' shape. In these cases considerable simplification
was achieved in comparison with previously published solutions.
A generalization of the obtained solution to the case of anisotropic media is discussed.
\end{abstract}

\pacs{68.65.Hb, 46.25.-y}

\maketitle

\section{Introduction}

Self-assembled quantum dots are three-dimensional inclusions of one
material in another one (a matrix). Usually there is a lattice
mismatch between materials of an inclusion and a matrix. The lattice
mismatch gives rise to a built-in inhomogeneous elastic strain which
in turn produces significant changes in the electronic band
structure. \cite{Bir_Pikus,Van_de_Walle} Therefore knowledge of the
strain distribution is of crucial importance for electronic
structure studies. Nearly all papers concerning electronic structure
calculations of quantum dots start with evaluation of elastic
strain. Especially important is the strain distribution for type-II
quantum dots where the confining potential for one type of carriers
is mainly due to the strain inhomogeneity. \cite{Yakimov}

There are a lot of theoretical works on the strain distribution in
quantum dot structures (for a review, see
Refs.~\onlinecite{Stangl2004_RMP,Maranganti2006_TCN}). In addition
to numerical calculations (using finite difference,
\cite{Grundmann1995,Pryor1998,Stier1999} finite element,
\cite{Christiansen1994,Noda1998} valence force field,
\cite{Cusack1996,Pryor1998,Stier1999,Nenashev2000,Kikuchi2001} and
molecular dynamics \cite{Daruka1999} methods), some analytical
techniques have been proposed. Most of them are based on the usage
of Green's functions, either in the real space
\cite{Faux1996,Downes1997_cuboid,Stoleru2002_pyramid} or in the
reciprocal space. \cite{Andreev1999} Some authors break the
inclusion into infinitely small ``bricks'' \cite{Downes1997_cuboid}
or into infinitely thin cuboids \cite{Glas2001} and then apply the
superposition principle. For ellipsoidal inclusions, Eshelby's
approach \cite{Eshelby1957} has proved to be effective. Also a
number of results obtained in thermoelasticity theory may be applied
to lattice-mismatched heterostructures, as pointed out in Ref.
\onlinecite{Davies1998}.

Different methods have their own merits and restrictions. In our
opinion, an ideal solution of the elastic inclusion problem has to
be analytical, to be expressed in terms of elementary functions and
written in closed form, to be applicable to a broad range of
inclusion shapes, and to take into account elastic anisotropy and
atomistic corrections. Analytical closed-form solutions have been
found for few cases of inclusion shapes: an ellipsoid,
\cite{Eshelby1957} a cuboid, \cite{Downes1997_cuboid} a pyramid,
\cite{Pearson2000_pyramid,Glas2001} and a variety of
quantum-wire-like structures. \cite{Faux1997_wire} Nozaki and
Taya~\cite{Nozaki2001_general_solution} have presented a general solution
for an arbitrary polyhedron, but it is extremely complicated. All these
solutions imply elastic isotropy and (except the case of ellipsoidal
inclusion) equal elastic constants of the two media.

The aim of our paper is to develop a novel approach to constructing
the solutions for the general case of a
polyhedral inclusion, and to propose a new insight into the
structure of a solution. We stress that the solution should have a
clear physical or geometrical meaning. Without having a clear
structure of a solution, it is hardly possible to develop its
generalization to anisotropic media and/or to inclusions with
elastic constants different from ones of the matrix.

This paper considers the following problem. There is an infinite
elastically isotropic medium (a matrix) with a finite
polyhedron-shaped inclusion. The crystal lattice of the inclusion
matches the lattice of the matrix without any defects. Elastic
moduli of the inclusion are assumed to be equal to ones of the
matrix, but the matrix and the inclusion have different lattice
constants. This produces an elastic strain in both the inclusion and
the matrix, and the task is to determine the strain tensor as a
function of coordinates, $\varepsilon_{\alpha\beta}({\mathbf r})$.
We neglect atomistic and nonlinearity effects, assuming that the
lattice mismatch is small, and lattice constants are small in
comparison with the inclusion size.

It is important to note that the strain distribution produced by an
inclusion in a \emph{semi-infinite} matrix may easily be calculated,
provided that the corresponding strain field in an \emph{infinite}
matrix is known. \cite{Davies2003_semi_inf} We will discuss it in
Section~\ref{sec:general}.

The rest of the paper is organized in the following way. In
Section~\ref{sec:method}, a new approach to evaluation the strain
distribution, based on an analogy between electrostatic and elastic
problems, is described. The solution for an arbitrary
polyhedron-shaped inclusion in an infinite matrix is presented and
discussed in Section~\ref{sec:general}. Then, in
Section~\ref{sec:pyr_and_hut}, this solution is applied to
pyramidal, truncated pyramidal, and ``hut-cluster'' inclusions.
Section~\ref{sec:anizo}
shows the possibility of generalization of our method to anisotropic media.
Section~\ref{sec:concl} contains the summary of the paper. The
Appendix is devoted to evaluation of solid angles that is important
for calculation of the strain.

\section{Electrostatic analogy}\label{sec:method}

The starting point of our investigation is a well-known analogy
between the elastic inclusion problem and the electrostatic problem
(Poisson equation). \cite{Davies1998} Namely, the \emph{displacement
vector} ${\mathbf u}({\mathbf r})$ induced by the inclusion is
proportional to the \emph{electric field} ${\mathbf F}({\mathbf r})$
that would appear if the inclusion were uniformly charged:
\begin{equation}\label{eq:u_ot_r}
{\mathbf u}({\mathbf r})=\frac{\varepsilon_0(1+\nu)}{4\pi(1-\nu)}
{\mathbf F}({\mathbf r})=\frac{\varepsilon_0(1+\nu)}{4\pi(1-\nu)}
\int\limits_V \frac{{\mathbf r}-{\mathbf r}'}{|{\mathbf r}-{\mathbf
r}'|^3}\, d{\mathbf r}',
\end{equation}
where $\varepsilon_0$ is the lattice mismatch
($\varepsilon_0=(a_\mathrm{inclusion}-a_\mathrm{matrix})/a_\mathrm{matrix}$,
$a$ being the lattice constant), $\nu$ is the Poisson ratio, $V$
denotes volume of the inclusion. For simplicity, in our auxiliary
electrostatic problem we take the charge density and the dielectric
constant equal to unity. Zero displacements correspond to positions
of atoms exactly in sites of the ideal lattice of the matrix.

Strain tensor is defined as
\begin{equation}\label{eq:eps_def}
\varepsilon_{\alpha\beta}(\mathbf r)=\frac{1}{2}\left(
\frac{\partial u_\alpha(\mathbf r)}{\partial x_\beta}+
\frac{\partial u_\beta(\mathbf r)}{\partial x_\alpha}\right)
-\varepsilon_0\,\delta_{\alpha\beta}\,\chi(\mathbf r),
\end{equation}
where $x_\alpha$ is $\alpha$-th component of the position vector
$\mathbf r$, $\delta_{\alpha\beta}$ is the Kroneker delta, and
$\chi(\mathbf r)$ is equal to 1 inside the inclusion and to 0
outside it.

Introducing an electrostatic potential
\begin{equation}\label{eq:fi_def}
\varphi(\mathbf r)=\int\limits_V \frac{d\mathbf r'}{|\mathbf
r-\mathbf r'|},
\end{equation}
we can express the field $\mathbf F(\mathbf r)$ as
\begin{equation}\label{eq:field}
\mathbf F(\mathbf r)=-\nabla\varphi(\mathbf r).
\end{equation}
Combination of Eqs.~(\ref{eq:u_ot_r}),~(\ref{eq:eps_def}) and (\ref{eq:field}) produces
\begin{equation}\label{eq:eps_ot_fi}
\varepsilon_{\alpha\beta}(\mathbf r)=
-\Lambda\,\frac{\partial^2\varphi(\mathbf r)}{\partial x_\alpha\partial x_\beta} -\varepsilon_0\,\delta_{\alpha\beta}\,\chi(\mathbf r),
\end{equation}
where $\Lambda=\varepsilon_0(1+\nu)/4\pi(1-\nu)$.

Now our aim is to evaluate second derivatives of the potential $\varphi(\mathbf r)$. For this purpose, we introduce three additional functions:
\newline 1) $\mathcal{F}_i(\mathbf r)$ --- an electrostatic potential of the uniformly charged (with unit surface density) $i$-th face of the inclusion surface;
\newline 2) $\Phi_k(\mathbf r)$ --- an electrostatic potential of the uniformly charged (with unit linear density) $k$-th edge of the inclusion surface;
\newline 3) $\Omega_i(\mathbf r)$ --- an electrostatic potential of the dipole layer uniformly spread over the $i$-th face with surface density of dipole moment equal to $\mathbf n^i$ --- the outward normal to the face.

The potential $\Phi(\mathbf r)$ of an uniformly charged edge is expressed as an integral $\int dl / \,|\mathbf r-\mathbf r'|$, where $dl$ is a linear element of the edge, and $\mathbf r'$ is a position vector of this linear element. Evaluation of this integral gives:
\begin{equation}\label{eq:Phi_k_def}
\Phi(\mathbf r) = \log\frac{r_1+r_2+L}{r_1+r_2-L}\,,
\end{equation}
where $r_1$ and $r_2$ are distances from the point $\mathbf r$ to ends of the edge, and $L$ is the edge length.

The potential $\Omega(\mathbf r)$ of a flat uniform dipole layer is known~\cite{Tamm} to be equal to the solid angle at which this layer is seen from the point $\mathbf r$, taken with positive sign if the positively charged side of the layer is seen from $\mathbf r$, and with negative sign otherwise. Thus below we will refer to the quantity $\Omega_i(\mathbf r)$ as to a solid angle subtended by the $i$-th face from the point $\mathbf r$.

\begin{figure}
\includegraphics[width=3in]{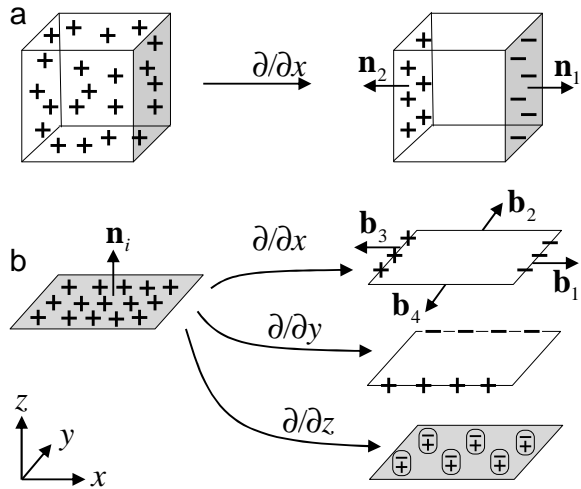}
\caption{\label{fig:cuboid} A sketch of modification of the ``charge distribution'' by taking derivatives of the potential: (a) transformation of a volume charge into a surface charge, (b) transformation of a surface charge into a linear charge and a dipole layer.}
\end{figure}

In order to find \emph{second} derivatives $\partial^2\varphi(\mathbf r)/\partial x_\alpha\partial x_\beta$, we note that the \emph{first} derivative can be expressed as a sum over faces of the inclusion surface:
\begin{equation}\label{eq:dfi}
\frac{\partial\varphi(\mathbf r)}{\partial\mathbf r} = -\sum_{\stackrel{\scriptstyle i}{\mathrm{(faces)}}} \mathbf{n}^i \, \mathcal{F}_i(\mathbf{r})
\end{equation}
(see Fig.~1a). Indeed, this derivative can be rewritten as
\begin{equation}\label{eq:dfiint}
\frac{\partial\varphi(\mathbf r)}{\partial\mathbf r} =
\frac{\partial}{\partial\mathbf r}\int\frac{\chi(\mathbf r')}{|\mathbf r-\mathbf r'|}\,d\mathbf r' =
\int \frac{\partial\chi(\mathbf r')}{\partial(\mathbf r')} \, \frac{d\mathbf r'}{|\mathbf r-\mathbf r'|} \,.
\end{equation}
The derivative $\partial\chi(\mathbf r')/\partial\mathbf r'$ plays the role of a ``charge density'' in Eq.~(\ref{eq:dfiint}). It vanishes everywhere except the surface of the inclusion. Near the $i$-th face of the surface, $\chi(\mathbf r)$ is equal to $-\theta(\mathbf{n}^i(\mathbf r-\mathbf r_i))$, where $\mathbf r_i$ is any point of this face, $\theta$ is the Heaviside function; consequently $\partial\chi(\mathbf r)/\partial\mathbf r= -\mathbf{n}^i\, \delta(\mathbf{n}^i(\mathbf r-\mathbf r_i))$, that corresponds to the ``surface charge density'' $-\mathbf{n}^i$ at the $i$-th face. Therefore, a contribution of the $i$-th face to $\partial\varphi(\mathbf r)/\partial\mathbf r$ is equal to $-\mathbf{n}^i \mathcal{F}_i$, according to Eq.~(\ref{eq:dfi}).

The next step is finding of derivatives of $\mathcal{F}_i(\mathbf r)$. Let, for simplicity, the $i$-th face lie in the plane $XY$, and its outward normal vector $\mathbf{n}^i$ be directed along the axis $Z$. To find derivatives $\partial\mathcal{F}_i/\partial x$ and $\partial\mathcal{F}_i/\partial y$, one can follow the same line of argumentation as at deriving Eq.~(\ref{eq:dfi}). The result is:
\begin{equation}\label{eq:dFxy}
\frac{\partial\mathcal{F}_i}{\partial x} = -\sum_k b^k_x \Phi_k, \quad
\frac{\partial\mathcal{F}_i}{\partial y} = -\sum_k b^k_y \Phi_k,
\end{equation}
where $k$ runs over edges surrounding the $i$-th face; $\mathbf{b}^k$ is a unit vector which is parallel to the $i$-th face, directed out of this face, and perpendicular to the $k$-th edge (see Fig.~1b). The last derivative, $\partial\mathcal{F}_i/\partial z$, transforms an uniformly charhed $i$-th face into a dipole layer with surface density of dipole momentum equal to $-\mathbf{n}^i$, as shown in Fig.~1b. Consequently,
\begin{equation}\label{eq:dFz}
\frac{\partial\mathcal{F}_i}{\partial z} = -n^i_z \Omega_i.
\end{equation}
Eqs.~(\ref{eq:dFxy}) and~(\ref{eq:dFz}) can be written together in a vector form, which is independent on orientation of a face with respect to co-ordinate axes:
\begin{equation}\label{eq:dFvector}
\frac{\partial\mathcal{F}_i(\mathbf r)}{\partial\mathbf r} =
-\mathbf n^i \Omega_i(\mathbf r) -\sum_k \mathbf b^k \Phi_k(\mathbf r).
\end{equation}

Using Eqs.~(\ref{eq:dfi}) and~(\ref{eq:dFvector}), one can express second derivatives of the potential $\varphi(\mathbf r)$ via solid angles $\Omega_i(\mathbf r)$ and potentials of charged edges $\Phi_k(\mathbf r)$:
\begin{eqnarray}\label{eq:d2firesult}
\nonumber \frac{\partial^2\varphi(\mathbf r)}{\partial x_\alpha\,\partial x_\beta} &=&
\sum_{\stackrel{\scriptstyle i}{\mathrm{(faces)}}} n^i_\alpha n^i_\beta \,\Omega_i(\mathbf{r}) \\
&+& \sum_{\stackrel{\scriptstyle k}{\mathrm{(edges)}}} (n^{k1}_\alpha b^{k1}_\beta+n^{k2}_\alpha b^{k2}_\beta) \,\Phi_k(\mathbf{r}),
\end{eqnarray}
where for each edge $k$ there are four unit vectors $\mathbf n^{k1}$, $\mathbf b^{k1}$, $\mathbf n^{k2}$, $\mathbf b^{k2}$, related to the two faces intersecting at this edge. With given normals $\mathbf n^{k1}$ and $\mathbf n^{k2}$, the vectors $\mathbf b^{k1}$ and $\mathbf b^{k2}$ can be found in the following way:
\begin{equation}\label{eq:b_def}
\mathbf b^{k1} =   \mathbf{n}^{k1} \times \mathbf{l}^{k}, \qquad
\mathbf b^{k2} = - \mathbf{n}^{k2} \times \mathbf{l}^{k},
\end{equation}
where $\mathbf{l}^{k}$ is a unit vector directed along edge $k$. From two possible directions of $\mathbf{l}^{k}$, one should choose the one going clockwise with respect to face $k1$ and, correspondingly, counter-clockwise with respect to $k2$, when seeing from the outside of the inclusion.

\section{General solution and its properties}\label{sec:general}

\subsection{The general solution}

In the previous Section, it was shown that the strain tensor
$\varepsilon_{\alpha\beta}(\mathbf{r})$ can be expressed (via
Eq.~(\ref{eq:eps_ot_fi})) in terms of second derivatives of some
auxiliary ``electrostatic potential'' $\varphi(\mathbf{r})$. In
turn, these second derivatives break
down into contributions of all faces
and edges of the inclusion surface (Eq.~(\ref{eq:d2firesult})).
Combining equations (\ref{eq:eps_ot_fi}) and
(\ref{eq:d2firesult}), we obtain the following expression for the
strain tensor:
\begin{eqnarray}\label{eq:main}
\nonumber\varepsilon_{\alpha\beta}(\mathbf{r})=&-&
\Lambda\sum_{\stackrel{\scriptstyle i}{\mathrm{(faces)}}}n_\alpha^in_\beta^i\Omega_i(\mathbf{r}) \\
&-&\Lambda\sum_{\stackrel{\scriptstyle
k}{\mathrm{(edges)}}}\gamma_{\alpha\beta}^k\Phi_k(\mathbf{r})
-\varepsilon_0\delta_{\alpha\beta}\chi(\mathbf{r}),
\end{eqnarray}
where $i$ runs over faces of the inclusion surface, and $k$ runs
over its edges.

In Eq.~(\ref{eq:main}), $\Omega_i(\mathbf{r})$ is a solid angle subtended by the $i$-th face from the point $\mathbf r$ (positive if the outer side of the face is seen from the point $\mathbf r$, and negative otherwise);
$\Phi_k(\mathbf{r})$ is the electrostatic potential of an uniformly charged $k$-th edge (with unit linear charge density) at the point $\mathbf r$;
$\chi(\mathbf{r})$ is equal to 1 inside the inclusion and to 0 outside it;
$\varepsilon_0$ is the relative lattice mismatch between the inclusion and the matrix;
the constant $\Lambda$ is equal to $\varepsilon_0(1+\nu)/4\pi(1-\nu)$, where $\nu$ is the Poisson ratio;
$\delta_{\alpha\beta}$ is the Kroneker delta;
$\mathbf n^i$ is a normal unit vector to the $i$-th face, directed outside the inclusion;
and a constant tensor $\gamma_{\alpha\beta}^{k}$ is equal to
\begin{equation}\label{eq:gamma_def}
\gamma_{\alpha\beta}^{k}= n_\alpha^{k1}b_\beta^{k1}+ n_\alpha^{k2}b_\beta^{k2}.
\end{equation}
In Eq.~(\ref{eq:gamma_def}),
$\mathbf n^{k1}$ and $\mathbf n^{k2}$ are normal unit
vectors (directed outside the inclusion) to the two faces which
intersect at the $k$-th edge;
$\mathbf b^{k1}$ is a unit vector perpendicular to the $k$-th
edge and to $\mathbf n^{k1}$, and directed out of the $k1$-th face;
analogously, $\mathbf b^{k2}$ is a unit vector perpendicular
to the $k$-th edge and to $\mathbf n^{k2}$, directed out of the
$k2$-th face (see Eq.~(\ref{eq:b_def})).

The tensor $\gamma_{\alpha\beta}^{k}$ is symmetrical, and it can be
written in an equivalent form
\[
\gamma_{\alpha\beta}^{k}=\left(
A_\alpha^kA_\beta^k-B_\alpha^kB_\beta^k \right)\sin\theta,
\]
where
\[
\mathbf A^k=\frac{\mathbf n^{k1}+\mathbf n^{k2}} {|\mathbf
n^{k1}+\mathbf n^{k2}|}, \quad\mathbf B^k=\frac{\mathbf
n^{k1}-\mathbf n^{k2}} {|\mathbf n^{k1}-\mathbf n^{k2}|},
\]
and $\theta$ is the internal dihedral angle between faces $k1$ and $k2$.

There is a simple expression (\ref{eq:Phi_k_def}) for potentials $\Phi_k(\mathbf{r})$ contributing into Eq.~(\ref{eq:main}).
Some closed-form expressions for solid angles $\Omega_i(\mathbf{r})$ are presented in the Appendix.

Equation~(\ref{eq:main}) is the main result of the present paper. It
gives a closed-form analytical expression for strain distribution in
and around a polyhedral inclusion buried into infinite isotropic
elastic medium.

With a known strain tensor, one can easily obtain the stress tensor
$\sigma_{\alpha\beta}$ via Hooke's law: \cite{Landau}
\begin{equation}\label{eq:Hookes_law}
\sigma_{\alpha\beta}(\mathbf{r})=\frac{E}{1+\nu}\left(
\varepsilon_{\alpha\beta}(\mathbf{r})+
\frac{\nu}{1-2\nu}\varepsilon_{\gamma\gamma}(\mathbf{r})\delta_{\alpha\beta}
\right),
\end{equation}
where $E$ is the Young modulus.

On the basis of Eq.~(\ref{eq:main}), the program code is written that can easily calculate the strain distribution produced by a lattice-mismatched inclusion in an infinite, isotropic matrix. Inclusion shape can be an arbitrary polyhedron. This program, named ``easystrain'', is freely available at {\tt http://easystrain.narod.ru}.

\subsection{Cuboidal inclusion}

If the inclusion has the form of cuboid with faces perpendicular to
the direction of the axes $x$, $y$ and $z$ (Fig.~\ref{fig:cuboid}),
then separate components of Eq.~(\ref{eq:main}) are simplified to
\begin{eqnarray*}
\varepsilon_{xx}(\mathbf{r}) &=&
-\Lambda\left(\Omega_1(\mathbf{r})+\Omega_2(\mathbf{r})\right)
-\varepsilon_0\chi(\mathbf{r}),\\
\varepsilon_{xy}(\mathbf{r}) &=&
-\Lambda\left(\Phi_1(\mathbf{r})-\Phi_2(\mathbf{r})
+\Phi_3(\mathbf{r})-\Phi_4(\mathbf{r})\right),
\end{eqnarray*}
and all the other components of strain tensor have a similar form.
So, the diagonal components $\varepsilon_{xx}$, $\varepsilon_{yy}$
and $\varepsilon_{zz}$ depend only on solid angles $\Omega_i$,
whereas off-diagonal components $\varepsilon_{xy}$,
$\varepsilon_{xz}$ and $\varepsilon_{yz}$ depend only on edge
contributions $\Phi_k$. This is in agreement with results of Downes,
Faux and O'Reilly. \cite{Downes1997_cuboid} These authors pointed
out that, in the case of cuboidal inclusion, solid angles subtended
by faces contribute into the stress tensor (and hence into the
strain tensor too). Our paper generalizes this observation to the
case of any polyhedral inclusion.

\begin{figure}
\includegraphics[width=2in]{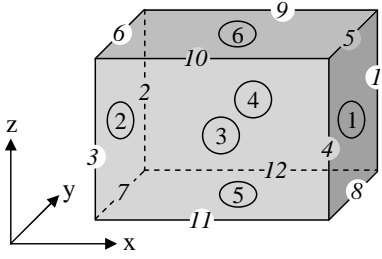}
\caption{\label{fig:cuboid} Cuboidal inclusion. Numbers in circles
refer to faces, italic numbers refer to edges.}
\end{figure}

\subsection{Hydrostatic strain}

Now we consider some simple properties of the solution
(\ref{eq:main}). These properties can be regarded as tests of
validity of the solution.

First, let us calculate the hydrostatic component of strain (that
is, the trace $\varepsilon_{\alpha\alpha}(\mathbf{r})$ of the strain
tensor). Taking into account that $n^i_\alpha n^i_\alpha=(\mathbf
n^i)^2=1$, $\gamma_{\alpha\alpha}^k= \mathbf n^{k1}\mathbf
b^{k1}+\mathbf n^{k2}\mathbf b^{k2}=0$, and
$\delta_{\alpha\alpha}=3$, we readily get from Eq.~(\ref{eq:main})
\[
\varepsilon_{\alpha\alpha}(\mathbf{r})=
-\Lambda\sum_i\Omega_i(\mathbf{r}) -3\varepsilon_0\chi(\mathbf{r}).
\]
The sum of solid angles, $\sum_i\Omega_i(\mathbf{r})$, vanishes for
any point $\mathbf{r}$ outside the inclusion. Indeed, all faces can
be divided into two groups with regard to the point $\mathbf{r}$:
1)~the ones whose outer sides are seen from the point $\mathbf{r}$,
2)~the ones whose inner sides are seen from $\mathbf{r}$. Net solid
angles subtended by the two groups are the same, but they contribute
to the sum $\sum_i\Omega_i(\mathbf{r})$ with opposite signs and
therefore cancel each other.

If the point $\mathbf{r}$ is inside the inclusion, all the faces
belong to the second group and the solid angle subtended by them
together are the full solid angle, $4\pi$. So,
$\sum_i\Omega_i(\mathbf{r})=-4\pi$. Combining both cases
($\mathbf{r}$ outside and inside the inclusion) we get
\begin{equation}\label{eq:sum_omega}
\sum_i\Omega_i(\mathbf{r})=-4\pi\chi(\mathbf{r}),
\end{equation}
and consequently
\begin{equation}\label{eq:eps_alpha_alpha}
\varepsilon_{\alpha\alpha}(\mathbf{r}) =
(4\pi\Lambda-3\varepsilon_0)\chi(\mathbf{r}) =
-2\varepsilon_0\frac{1-2\nu}{1-\nu}\chi(\mathbf{r}).
\end{equation}
We have come to the well-known result that the hydrostatic strain is
zero outside the inclusion and is constant inside it.
\cite{Davies1998}

\subsection{Strain discontinuities at faces}

Then, it is easy to examine the behavior of the strain at the
inclusion surface, starting from Eq.~(\ref{eq:main}). When the point
$\mathbf r$, moving from outside toward inside, crosses a face of
the inclusion, the strain changes stepwise. The discontinuity of the
strain, $\Delta\varepsilon_{\alpha\beta}\equiv
\varepsilon_{\alpha\beta}^\mathrm{inside}
-\varepsilon_{\alpha\beta}^\mathrm{outside}$, can be written as
\[
\Delta\varepsilon_{\alpha\beta}=
-\Lambda\sum_in_\alpha^in_\beta^i\Delta\Omega_i
-\Lambda\sum_k\gamma_{\alpha\beta}^k\Delta\Phi_k
-\varepsilon_0\delta_{\alpha\beta}.
\]
There $\Delta\Omega_i$ and $\Delta\Phi_k$ denote discontinuities of
$\Omega_i$ and $\Phi_k$. In fact, edge contributions $\Phi_k$ have
no discontinuities at the face, and only one of $\Omega_i$ has a
discontinuity---namely, the $\Omega_i$ related to the face under
consideration. For this face, $\Delta\Omega_i=-4\pi$, because the
solid angle $|\Omega_i|$ reaches $2\pi$ at the face, and the value
$\Omega_i$ changes the sign from positive to negative. So,
\[
\Delta\varepsilon_{\alpha\beta}= 4\pi\Lambda n_\alpha n_\beta
-\varepsilon_0\delta_{\alpha\beta}= \varepsilon_0\left(
\frac{1+\nu}{1-\nu}\,n_\alpha n_\beta-\delta_{\alpha\beta} \right),
\]
with a unit vector $\mathbf n$ perpendicular to the face. It is
convenient to consider the strain tensor with respect to the
coordinate axes $\xi$, $\eta$, $\zeta$ connected to the face: axes
$\xi$ and $\eta$ parallel to the face, and the axis $\zeta$
perpendicular to it. Therefore $n_\xi=n_\eta=0$, $n_\zeta=1$, and
$\Delta\varepsilon_{\alpha\beta}$ takes the following form:
\begin{equation}\label{eq:Delta_eps}
\Delta\varepsilon_{\xi\xi}=\Delta\varepsilon_{\eta\eta}=-\varepsilon_0,
\quad\Delta\varepsilon_{\zeta\zeta}=\frac{2\varepsilon_0\nu}{1-\nu},
\end{equation}
off-diagonal components $\Delta\varepsilon_{\xi\eta}$,
$\Delta\varepsilon_{\xi\zeta}$, $\Delta\varepsilon_{\eta\zeta}$ are
zero. According to Hooke's law (\ref{eq:Hookes_law}),
discontinuities of the stress tensor, $\Delta\sigma_{\alpha\beta}$,
are
\begin{equation}\label{eq:Delta_sigma}
\Delta\sigma_{\xi\xi}=\Delta\sigma_{\eta\eta}=-\frac{E}{1-\nu},
\quad\Delta\sigma_{\zeta\zeta}=0,
\end{equation}
and again off-diagonal components are zero.

Equations (\ref{eq:Delta_eps}) and (\ref{eq:Delta_sigma}) are in
accordance to the boundary conditions at the interface:
$\Delta\varepsilon_{\xi\xi}=\Delta\varepsilon_{\eta\eta}=-\varepsilon_0$,
$\Delta\varepsilon_{\xi\eta}=0$ (continuity of displacement field
$\mathbf u(\mathbf r)$),
$\Delta\sigma_{\xi\zeta}=\Delta\sigma_{\eta\zeta}=\Delta\sigma_{\zeta\zeta}=0$
(balance of elastic forces at the interface).

\begin{figure}
\includegraphics[width=2.8in]{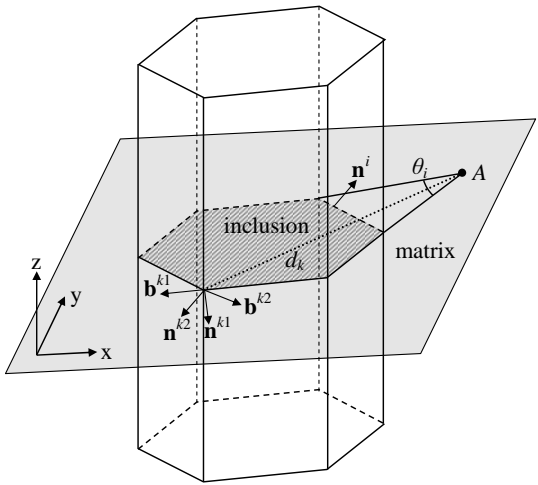}
\caption{\label{fig:wire} Quantum wire inclusion. Point $A$ is the
point where the strain is to be obtained. $\theta_i$ and $d_k$ are
the angle and distance contributing to Eq.~(\ref{eq:main_2D}).}
\end{figure}

\subsection{Quantum-wire inclusion}

Next, we consider the strain distribution in a quantum-wire-like
inclusion and its surrounding (Fig.~\ref{fig:wire}). Such an
inclusion is a prism, the top and bottom of which go to infinity.
For simplicity, let all side faces and edges be parallel to the axis
$z$. So the strain is independent on~$z$.

To obtain the strain distribution, one may start from
Eq.~(\ref{eq:main}) for a prism of finite height, and then go to the
limit of infinitely large vertical dimension. In this limit,
contributions of base and bottom faces, as well as of edges
adjoining to these faces, vanish. For each side face, the solid
angle $\Omega_i$ reduces to a doubled plane angle $\theta_i$
(Fig.~\ref{fig:wire}) subtended by the cross-section of this face by
a plane parallel to axes $x$, $y$: $\Omega_i=2\theta_i$. Edge
contributions $\Phi_k$ reduce to simple logarithmic expressions:
\[
\Phi_k=-2\log d_i+\mathrm{const},
\]
where $d_k$ is the distance to the $k$-th edge
(Fig.~\ref{fig:wire}). The constants in these expressions are
infinitely large, but they cancel each other being substituted into
Eq.~(\ref{eq:main}).

As a result, we come to the following expression for the strain in a
quantum-wire-like inclusion:
\begin{eqnarray}\label{eq:main_2D}
\nonumber\varepsilon_{\alpha\beta}(\mathbf r)=&-&2\Lambda\sum_i
n_\alpha^i n_\beta^i \theta_i \\
&+& 2\Lambda\sum_k\gamma_{\alpha\beta}^k\log d_k -
\varepsilon_0\delta_{\alpha\beta}\chi.
\end{eqnarray}
There the indices $i$ and $k$ run over all side faces and edges,
correspondingly; $\theta_i$ is the plane angle subtended by the
cross section of $i$-th face by the plane passing through the point
$\mathbf r$ parallel to the axes $x$ and $y$ (positive if the outer
side of the face is seen from the point $\mathbf r$, and negative
otherwise); $d_k$ is the distance from the point $\mathbf r$ to
$k$-th edge. All the rest notations are the same as in Eq.~(\ref{eq:main}).
As the $xz$-, $yz$- and
$zz$-components of the tensors $n^i_\alpha n^i_\beta$ and
$\gamma_{\alpha\beta}^k$ are zero, the corresponding components of
strain tensor are independent on $\theta_i$ and $d_k$:
\[
\varepsilon_{xz}=\varepsilon_{yz}=0, \quad
\varepsilon_{zz}=-\varepsilon_0\chi.
\]

Equation~(\ref{eq:main_2D}) is an equivalent, but more simple and
compact, form of the solution obtained by Faux, Downes and O'Reilly.
\cite{Faux1997_wire}

\subsection{Semi-infinite matrix}

Finally we discuss the strain distribution in a semi-infinite
matrix. Davies~\cite{Davies2003_semi_inf} proposed a method of
reducing the elastic inclusion problem in a \emph{semi-infinite}
matrix to the corresponding problem in an \emph{infinite} matrix.
For convenience, we reproduce there the results of Davies's work.
\cite{Davies2003_semi_inf}

Let an inclusion be buried in a semi-infinite matrix that fill a
half-space $z>z_s$, or $z<z_s$, with a free surface in the plane
$z=z_s$. Isotropic linear elasticity is assumed, and elastic moduli
of the matrix and the inclusion are the same. To calculate the
strain distribution $\tilde{\varepsilon}_{\alpha\beta}(\mathbf{r})$
in this system, one can previously find an analogous strain
distribution $\varepsilon_{\alpha\beta}(\mathbf{r})$ in a system
consisting of the same inclusion in an infinite matrix. It can be
found by Eq.~(\ref{eq:main}), for example. Then, components of
$\tilde{\varepsilon}_{\alpha\beta}(\mathbf{r})$ are expressed via
components of $\varepsilon_{\alpha\beta}(\mathbf{r})$,
$\varepsilon_{\alpha\beta}(\mathbf{r}_2)$ and their derivatives
$\partial\varepsilon_{\alpha\beta}/\partial z(\mathbf{r}_2)$, where
the point $\mathbf{r}_2$ is a ``mirror image'' of the point
$\mathbf{r}$ with respect to the surface:
\begin{eqnarray*}
\mathbf r &=& (x,y,z),\quad \mathbf r_2 = (x,y,2z_s-z),\\
\tilde{\varepsilon}_{xx}(\mathbf{r}) &=&
\varepsilon_{xx}(\mathbf{r})+(3-4\nu)\varepsilon_{xx}(\mathbf{r}_2)
+2(z-z_s)\frac{\partial\varepsilon_{xx}}{\partial z}(\mathbf{r}_2),\\
\tilde{\varepsilon}_{yy}(\mathbf{r}) &=&
\varepsilon_{yy}(\mathbf{r})+(3-4\nu)\varepsilon_{yy}(\mathbf{r}_2)
+2(z-z_s)\frac{\partial\varepsilon_{yy}}{\partial z}(\mathbf{r}_2),\\
\tilde{\varepsilon}_{zz}(\mathbf{r}) &=&
\varepsilon_{zz}(\mathbf{r})-(1-4\nu)\varepsilon_{zz}(\mathbf{r}_2)
+2(z-z_s)\frac{\partial\varepsilon_{zz}}{\partial z}(\mathbf{r}_2),\\
\tilde{\varepsilon}_{xy}(\mathbf{r}) &=&
\varepsilon_{xy}(\mathbf{r})+(3-4\nu)\varepsilon_{xy}(\mathbf{r}_2)
+2(z-z_s)\frac{\partial\varepsilon_{xy}}{\partial z}(\mathbf{r}_2),\\
\tilde{\varepsilon}_{xz}(\mathbf{r}) &=&
\varepsilon_{xz}(\mathbf{r})-\varepsilon_{xz}(\mathbf{r}_2)
-2(z-z_s)\frac{\partial\varepsilon_{xz}}{\partial z}(\mathbf{r}_2),\\
\tilde{\varepsilon}_{yz}(\mathbf{r}) &=&
\varepsilon_{yz}(\mathbf{r})-\varepsilon_{yz}(\mathbf{r}_2)
-2(z-z_s)\frac{\partial\varepsilon_{yz}}{\partial z}(\mathbf{r}_2).
\end{eqnarray*}

If the inclusion is a polyhedron, Eq.~(\ref{eq:main}) provides an
analytical expression for the strain tensor
$\varepsilon_{\alpha\beta}$. Therefore its derivative
$\partial\varepsilon_{\alpha\beta}/\partial z$ can be evaluated
analytically as a combination of derivatives of solid angles
$\Omega_i$ and values $\Phi_k$. It is important to note that
derivatives of $\Omega_i$ can be expressed via derivatives of
$\Phi_k$:
\begin{equation}\label{eq:dOmega_via_dPhi}
\frac{\partial\Omega_i(\mathbf r)}{\partial x_\alpha}=
\sum_k\left( b^{ki}_\alpha n^i_\beta - b^{ki}_\beta n^i_\alpha \right)
\frac{\partial\Phi_k(\mathbf{r})}{\partial x_\beta},
\end{equation}
where summation is over all the edges adjoining to the $i$-th face;
and $\mathbf{b}^{ki}$ is the one of unit vectors $\mathbf{b}^{k1}$,
$\mathbf{b}^{k2}$, which is perpendicular to $\mathbf{n}^i$. Eq.~(\ref{eq:dOmega_via_dPhi}) may be useful
since analytical expressions for solid angles are rather complicated
in comparison with the expression~(\ref{eq:Phi_k_def}) for values
$\Phi_k$.

\section{Application to pyramidal and hut-cluster inclusions}\label{sec:pyr_and_hut}

Among all polyhedrons, the three ones appear most often as
geometrical models of quantum dots. These are square-based pyramid
(Fig.~\ref{fig:pyr_hat}a), truncated square-based pyramid
(Fig.~\ref{fig:pyr_hat}b), and so-called ``hut-cluster''
(Fig.~\ref{fig:pyr_hat}c). In this Section, we apply the general
expression (\ref{eq:main}) to the specific cases of pyramidal and
hut-cluster inclusions. The case of truncated pyramid does not
demand a special consideration, because it is easy to obtain
solution for truncated pyramid, provided that the solution for
pyramid has yet been obtained (see below).

\subsection{Pyramid}

\begin{figure}
\includegraphics[width=3in]{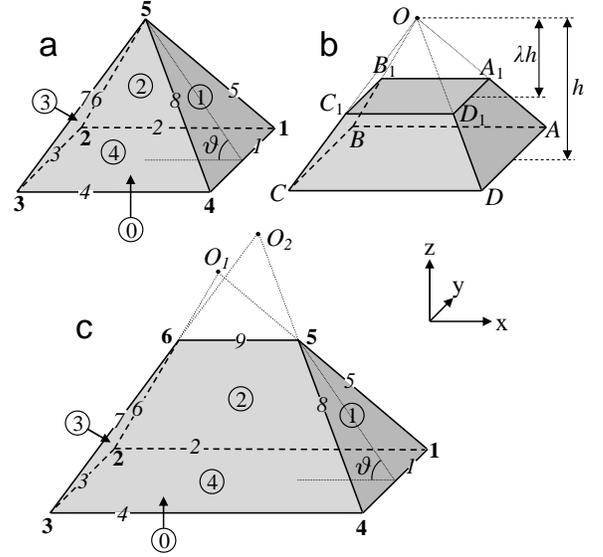}
\caption{\label{fig:pyr_hat} Inclusions of most common shapes: (a) a
pyramid with square base; (b) a truncated pyramid with square base;
(c) a ``hut-cluster''. Numbers in circles refer to faces (the base
has the number 0), italic numbers refer to edges, and bold
numbers---to vertices.}
\end{figure}

With the numbering scheme of Fig.~\ref{fig:pyr_hat}a, we get the
following expressions for tensors $n^i_\alpha n^i_\beta$ and
$\gamma_{\alpha\beta}^k$:
\begin{eqnarray*}
n_\alpha^0n_\beta^0&=&\left\{0,\,0,\,1,\,0,\,0,\,0\right\},\\
n_\alpha^1n_\beta^1&=&\left\{s^2,\,0,\,c^2,\,0,\,-sc,\,0\right\},\\
n_\alpha^2n_\beta^2&=&\left\{0,\,s^2,\,c^2,\,0,\,0,\,-sc\right\},\\
n_\alpha^3n_\beta^3&=&\left\{s^2,\,0,\,c^2,\,0,\,sc,\,0\right\},\\
n_\alpha^4n_\beta^4&=&\left\{0,\,s^2,\,c^2,\,0,\,0,\,sc\right\},\\
\gamma_{\alpha\beta}^1&=&s\times\left\{c,\,0,\,-c,\,0,\,-s,\,0\right\},\\
\gamma_{\alpha\beta}^2&=&s\times\left\{0,\,c,\,-c,\,0,\,0,\,-s\right\},\\
\gamma_{\alpha\beta}^3&=&s\times\left\{c,\,0,\,-c,\,0,\,s,\,0\right\},\\
\gamma_{\alpha\beta}^4&=&s\times\left\{0,\,c,\,-c,\,0,\,0,\,s\right\},\\
\gamma_{\alpha\beta}^5&=&s/\sqrt{1+c^2}\times\left\{-c^2,\,-c^2,\,2c^2,\,1,\,sc,\,sc\right\},\\
\gamma_{\alpha\beta}^6&=&s/\sqrt{1+c^2}\times\left\{-c^2,\,-c^2,\,2c^2,\,-1,\,-sc,\,sc\right\},\\
\gamma_{\alpha\beta}^7&=&s/\sqrt{1+c^2}\times\left\{-c^2,\,-c^2,\,2c^2,\,1,\,-sc,\,-sc\right\},\\
\gamma_{\alpha\beta}^8&=&s/\sqrt{1+c^2}\times\left\{-c^2,\,-c^2,\,2c^2,\,-1,\,sc,\,-sc\right\},\\
\end{eqnarray*}
where $s=\sin\vartheta$, $c=\cos\vartheta$, and $\vartheta$ is a
dihedral angle between the pyramid base and any of its side face.
The tensor components are listed in braces in the following order:
$xx$, $yy$, $zz$, $xy$, $xz$, $yz$.

It is worth to note the following property of the set of tensors
$\gamma_{\alpha\beta}^k$:
\[
\sum_k \gamma_{\alpha\beta}^k L_k=0,
\]
where $L_k$ is the length of the $k$-th edge. This property comes
from a requirement that all terms proportional to $r^{-1}$ in
Eq.~(\ref{eq:main}) must cancel each other at $r\rightarrow\infty$.
It may serve as a useful test of correctness of the results.

These values of $n^i_\alpha n^i_\beta$ and $\gamma_{\alpha\beta}^k$,
together with Eq.~(\ref{eq:main}), give the expression for the
strain distribution in a pyramidal inclusion and its surrounding:
\begin{widetext}
\begin{eqnarray}\label{eq:pyramid_solution}
\nonumber\varepsilon_{xx} &=& -s^2\Lambda(\Omega_1+\Omega_3)
-sc\Lambda(\Phi_1+\Phi_3)
+\textstyle\frac{sc^2\Lambda}{\sqrt{1+c^2}}\Phi_{5-8} -\varepsilon_0\chi,\\
\nonumber\varepsilon_{yy} &=& -s^2\Lambda(\Omega_2+\Omega_4)
-sc\Lambda(\Phi_2+\Phi_4)
+\textstyle\frac{sc^2\Lambda}{\sqrt{1+c^2}}\Phi_{5-8} -\varepsilon_0\chi,\\
\varepsilon_{zz} &=& -\Lambda\Omega_0 -c^2\Lambda\Omega_{1-4}
+sc\Lambda\Phi_{1-4}
-\textstyle\frac{2sc^2\Lambda}{\sqrt{1+c^2}}\Phi_{5-8} -\varepsilon_0\chi,\\
\nonumber\varepsilon_{xy} &=&
-\textstyle\frac{s\Lambda}{\sqrt{1+c^2}}(\Phi_5-\Phi_6+\Phi_7-\Phi_8),\\
\nonumber\varepsilon_{xz} &=& sc\Lambda(\Omega_1-\Omega_3)
+s^2\Lambda(\Phi_1-\Phi_3)
-\textstyle\frac{s^2c\Lambda}{\sqrt{1+c^2}}(\Phi_5-\Phi_6-\Phi_7+\Phi_8),\\
\nonumber\varepsilon_{yz} &=& sc\Lambda(\Omega_2-\Omega_4)
+s^2\Lambda(\Phi_2-\Phi_4)
-\textstyle\frac{s^2c\Lambda}{\sqrt{1+c^2}}(\Phi_5+\Phi_6-\Phi_7-\Phi_8).
\end{eqnarray}
\end{widetext}
There we use a shorthand notation:
$\Omega_{1-4}=\Omega_1+\Omega_2+\Omega_3+\Omega_4$, and so on. Note
that, using Eq.~(\ref{eq:sum_omega}), one can simplify the
expression for $\varepsilon_{zz}$ to the following one:
\begin{eqnarray}\label{eq:pyramid_epszz_simplified}
\nonumber\varepsilon_{zz} &=& -s^2\Lambda\Omega_0 +sc\Lambda\Phi_{1-4}\\
&&-\textstyle\frac{2sc^2\Lambda}{\sqrt{1+c^2}}\Phi_{5-8}
+\varepsilon_0(c^2\textstyle\frac{1+\nu}{1-\nu}-1)\chi.
\end{eqnarray}
Eq.~(\ref{eq:solid_pyramid}) in the Appendix provides analytical
expressions for solid angles $\Omega_0...\Omega_4$ in the pyramid.

\subsection{Truncated pyramid}

To get the solution for the \emph{truncated} pyramid,
$\varepsilon_{\alpha\beta}^\mathrm{(trunc)}$, the easiest way is to
start from the solution for a pyramid,
$\varepsilon_{\alpha\beta}^\mathrm{(pyr)}$, and apply the
superposition principle. The full pyramid, $OABCD$, consists of a
truncated one, $ABCDA_1B_1C_1D_1$, and a smaller one,
$OA_1B_1C_1D_1$ (Fig.~\ref{fig:pyr_hat}b). According to the
superposition principle,
\[
\varepsilon_{\alpha\beta}^\mathrm{(pyr)}(\mathbf{r}) =
\varepsilon_{\alpha\beta}^\mathrm{(trunc)}(\mathbf{r}) +
\varepsilon_{\alpha\beta}^\mathrm{(small)}(\mathbf{r}),
\]
where $\varepsilon_{\alpha\beta}^\mathrm{(pyr)}$,
$\varepsilon_{\alpha\beta}^\mathrm{(trunc)}$ and
$\varepsilon_{\alpha\beta}^\mathrm{(small)}$ refer to figures
$OABCD$, $ABCDA_1B_1C_1D_1$ and $OA_1B_1C_1D_1$, correspondingly.
Then, it is well known that, in the framework of the continual
elasticity theory, similar inclusions produce similar strain fields.
As pyramids $OABCD$ and $OA_1B_1C_1D_1$ are similar, there is a
relation between $\varepsilon_{\alpha\beta}^\mathrm{(pyr)}$ and
$\varepsilon_{\alpha\beta}^\mathrm{(small)}$:
\[
\varepsilon_{\alpha\beta}^\mathrm{(pyr)}(\mathbf{r}_O+\mathbf{r}) =
\varepsilon_{\alpha\beta}^\mathrm{(small)}(\mathbf{r}_O+\lambda\mathbf{r}),
\]
where $\mathbf r_O$ is a position vector of the apex $O$, $\lambda$
is a truncation parameter (a ratio of sizes of the two pyramids, see
Fig.~\ref{fig:pyr_hat}b). So
$\varepsilon_{\alpha\beta}^\mathrm{(trunc)}$ can be expressed in
terms of $\varepsilon_{\alpha\beta}^\mathrm{(pyr)}$:
\begin{equation}\label{eq:truncated_solution}
\varepsilon_{\alpha\beta}^\mathrm{(trunc)}(\mathbf{r})=
\varepsilon_{\alpha\beta}^\mathrm{(pyr)}(\mathbf{r})
-\varepsilon_{\alpha\beta}^\mathrm{(pyr)}(\frac{\mathbf{r}-\mathbf{r}_O}{\lambda}+\mathbf{r}_O).
\end{equation}

This solution was compared numerically with the solution published
in Ref.~\onlinecite{Pearson2000_pyramid}. We reproduce all strain
profiles presented in that paper, except the component
$\varepsilon_{xz}$ in Fig.~11 of
Ref.~\onlinecite{Pearson2000_pyramid}, where the absolute value
coincides with our results, but the sign was opposite. We believe
that the sign of $\varepsilon_{xz}$ in
Ref.~\onlinecite{Pearson2000_pyramid} is erroneous, because it leads
to an incorrect behavior of the strain at $r\rightarrow\infty$.
Indeed, the multipole expansion, being applied to
Eq.~(\ref{eq:u_ot_r}), gives for large $r$
\[
\varepsilon_{\alpha\beta}(\mathbf{r}+\mathbf{r}_c) =
\frac{\varepsilon_0V(1+\nu)}{4\pi(1-\nu)}\;
\frac{\delta_{\alpha\beta}-3r_\alpha r_\beta/r^2}{r^3} + O(r^{-5}),
\]
where $V$ is a volume of the inclusion, and $\mathbf{r}_c$ is a
position vector of its center of mass. For a pyramid, $x_c=y_c=0$.
This expression shows that, at fixed positive $x$ and $y$,
$\varepsilon_{xz}$ must be negative when $z\rightarrow+\infty$ and
positive when $z\rightarrow-\infty$. This predicted behavior of
$\varepsilon_{xz}$ disagrees with Fig.~11 of
Ref.~\onlinecite{Pearson2000_pyramid}, but agrees with our
calculations.

\subsection{Hut-cluster}

Finally, we consider the hut-cluster. The hut-cluster is a figure
that consists of the base (a parallelogram) and four side faces.
Slope angles of all the side faces are the same. Therefore
orientations of faces and edges of the hut-cluster are the same as
of pyramid, except the top (9th) edge. So, the solution for the
hut-cluster is very similar to the one for the pyramid. The only
difference is the addition of the contribution of 9th edge. Of
course, the values of solid angles $\Omega_0...\Omega_4$ and of edge
contributions $\Phi_1...\Phi_8$ in the hut-cluster are not the same
as in the pyramid. Analytical expressions for solid angles in the
hut-cluster are given by Eq.~(\ref{eq:solid_hut}) in the Appendix.

To get the solution for the hut-cluster from
Eq.~(\ref{eq:pyramid_solution}), it is sufficient to add the term
$2sc\Lambda\Phi_9$ to $\varepsilon_{yy}$, and to add the term
$-2sc\Lambda\Phi_9$ to $\varepsilon_{zz}$. This demonstrates the
flexibility of the general solution (\ref{eq:main}). This is a
property that is not inherent in previous particular solutions.
\cite{Pearson2000_pyramid,Glas2001,Stoleru2002_pyramid}

An analytical solution for a hut-cluster was first obtained by
Glas\cite{Glas2001} as a special case of a more general answer for a
truncated pyramid with rectangular bottom and top faces. Our method
provides a much more simple solution.

\subsection{Strain profiles along axes of symmetry}

If the point $\mathbf r$ lies at the four-fold axis of symmetry of
the pyramid, expressions (\ref{eq:pyramid_solution}) are simplified
greatly, because all values
$\Omega_1(\mathbf{r})...\Omega_4(\mathbf{r})$ are the same, values
$\Phi_1(\mathbf{r})...\Phi_4(\mathbf{r})$ are the same, and
$\Phi_5(\mathbf{r})...\Phi_8(\mathbf{r})$ are also the same.
Moreover, it is sufficient to evaluate only the $zz$-component of
the strain, because non-diagonal components $\varepsilon_{xy}$,
$\varepsilon_{xz}$, $\varepsilon_{yz}$ are zero, and other diagonal
components $\varepsilon_{xx}$ and $\varepsilon_{yy}$ can be
expressed via $\varepsilon_{zz}$ using
Eq.~(\ref{eq:eps_alpha_alpha}):
\begin{equation}\label{eq:epsxx_from_epszz}
\varepsilon_{xx}=\varepsilon_{yy}=
\frac12(\varepsilon_{\alpha\alpha}-\varepsilon_{zz})=
-\varepsilon_0\frac{1-2\nu}{1-\nu}\chi -\frac12\varepsilon_{zz}.
\end{equation}
It is convenient to use Eq.~(\ref{eq:pyramid_epszz_simplified}) for
evaluation the component $\varepsilon_{zz}$. Let us put the origin
of coordinate system to the center of the pyramid base. So,
coordinates of a point at the axis of symmetry are $(0,0,z)$. Let
$h$ be a pyramid height, $a$ be a length of the pyramid base,
$l=\sqrt{a^2/2+h^2}$ be a length of a side edge, and $r$ be a
distance between the point $(0,0,z)$ and any vertex of the pyramid
base:
\[
r=\sqrt{a^2/2+z^2}.
\]
According to Eq.~(\ref{eq:solid_pyramid}), the solid angle
$\Omega_0$ is equal to
\[
\Omega_0= -4\arctan\frac{a^2}{4zr}.
\]
Then,
\begin{eqnarray*}
\Phi_1=\Phi_2=\Phi_3=\Phi_4&=&\log\frac{2r+a}{2r-a},\\
\Phi_5=\Phi_6=\Phi_7=\Phi_8&=&\log\frac{|z-h|+r+l}{|z-h|+r-l}.
\end{eqnarray*}
Substituting all these quantities into
Eq.~(\ref{eq:pyramid_epszz_simplified}), we get
\begin{eqnarray}\label{eq:profile_pyramid}
\nonumber \varepsilon_{zz}(0,0,z) &=&
\frac{8ah\Lambda}{a^2+4h^2}\left(
\frac{2h}{a}\arctan\frac{a^2}{4zr} +\log\frac{2r+a}{2r-a} \right.\\
&-& \left.\frac{a}{l}\log\frac{|z-h|+r+l}{|z-h|+r-l} \right)
+\tilde{\Lambda}\chi.
\end{eqnarray}
Here we expressed quantities $s$ and $c$ in terms of $a$, $h$, and
$l$; $\Lambda=\varepsilon_0(1+\nu)/4\pi(1-\nu)$; $\chi=1$ if
$z\in(0;h)$ and $\chi=0$ otherwise; the constant $\tilde{\Lambda}$
is the coefficient at $\chi$ in
Eq.~(\ref{eq:pyramid_epszz_simplified}):
\[
\tilde{\Lambda} = \varepsilon_0(c^2\textstyle\frac{1+\nu}{1-\nu}-1)
\equiv
\varepsilon_0\left(\frac{a^2(1+\nu)}{(a^2+4h^2)(1-\nu)}-1\right).
\]

For the truncated pyramid, Eq.~(\ref{eq:truncated_solution})
together with Eq.~(\ref{eq:profile_pyramid}) give
\begin{eqnarray}\label{eq:profile_truncated}
\nonumber &&\!\!\!\!\!\!\!\!\varepsilon_{zz}(0,0,z) =
\frac{8ah\Lambda}{a^2+4h^2}\left( \frac{2h}{a}\arctan\frac{a^2}{4zr}
-\frac{2h}{a}\arctan\frac{a^2}{4\tilde{z}\tilde{r}}\right.\\
&&\!\!\!\!\!\!\!\!+\left.\log\!\frac{(2r\!+\!a)(2\tilde{r}\!-\!a)}{(2r\!-\!a)(2\tilde{r}\!+\!a)}
-\frac{a}{l}\log\!\frac{\lambda\tilde{r}\!+\!r\!+\!(1\!-\!\lambda)l}{\lambda\tilde{r}\!+\!r\!-\!(1\!-\!\lambda)l}
\right) \!+\!\tilde{\Lambda}\chi.
\end{eqnarray}
In Eq.~(\ref{eq:profile_truncated}), $\chi=1$ if
$z\in(0;(1-\lambda)h)$, and $\chi=0$ otherwise;
$\tilde{z}=(z-h)/\lambda+h$; $\tilde{r}=\sqrt{a^2/2+\tilde{z}^2}$ is
a distance between the point $(0,0,\tilde{z})$ and any vertex of the
base; $l=\sqrt{a^2/2+h^2}$.

In a similar manner, one can get the strain profile along the axis
of symmetry of the hut-cluster. As there is only two-fold axis in
the hut-cluster, the components $\varepsilon_{xx}$ and
$\varepsilon_{yy}$ are no longer the same. Therefore we cannot use
Eq.~(\ref{eq:epsxx_from_epszz}) to extract $\varepsilon_{xx}$ and
$\varepsilon_{yy}$ from $\varepsilon_{zz}$. Instead, we should find
$\varepsilon_{zz}$ and $\varepsilon_{xx}$ independently, and then
extract $\varepsilon_{yy}$ by means of
Eq.~(\ref{eq:eps_alpha_alpha}):
\[
\varepsilon_{yy}=
\varepsilon_{\alpha\alpha}-\varepsilon_{xx}-\varepsilon_{zz}=
-2\varepsilon_0\frac{1-2\nu}{1-\nu}\chi
-\varepsilon_{xx}-\varepsilon_{zz}.
\]
We chose the center of the cluster base as an origin of the
coordinate system. Let $h$ be a cluster height, $a$ and $b$ be the
smaller and the bigger edge lengths of the base, correspondingly.
Then, values of $\varepsilon_{zz}$ and $\varepsilon_{xx}$ at the
axis of symmetry are
\begin{widetext}
\begin{equation}\label{eq:profile_hut_epszz}%\footnotesize
\varepsilon_{zz}(0,0,z) \!=\! \frac{8ah\Lambda}{a^2\!+\!4h^2}\left(
\frac{2h}{a}\arctan\frac{ab}{4zr_1}
+\frac12\log\frac{(2r_1\!+\!a)(2r_1\!+\!b)}{(2r_1\!-\!a)(2r_1\!-\!b)}
-\frac{a}{l}\log\frac{r_5\!+\!r_1\!+\!l}{r_5\!+\!r_1\!-\!l}
-\frac12\log\frac{2r_5\!+\!(b\!-\!a)}{2r_5\!-\!(b\!-\!a)} \right)
+\tilde{\Lambda}\chi,
\end{equation}
\begin{equation}\label{eq:profile_hut_epsxx}%\footnotesize
\varepsilon_{xx}(0,0,z) \!=\!
\frac{8ah\Lambda}{a^2\!+\!4h^2}\!\left(\!
-\frac{2h}{a}\arctan\frac{a^2z-abh}{[b(b\!-\!a)\!+\!4z(z\!-\!h)]r_1\!+\!(b^2\!+\!4z^2)r_5}
-\!\frac12\log\frac{2r_1\!+\!a}{2r_1\!-\!a}
+\!\frac{a}{2l}\log\frac{r_5\!+\!r_1\!+\!l}{r_5\!+\!r_1\!-\!l}
\right) -\varepsilon_0\chi,
\end{equation}
\end{widetext}
where $r_1(z)=\sqrt{a^2/4+b^2/4+z^2}$ is a distance from the point
$(0,0,z)$ to the first vertex of the hut-cluster;
$r_5(z)=\sqrt{(b-a)^2/4+(z-h)^2}$ is a distance from the point
$(0,0,z)$ to the fifth vertex; $l=\sqrt{a^2/2+h^2}$ is a length of
each side edge; $\chi=1$ if $z\in(0;h)$ and $\chi=0$ otherwise. The
last log term in Eq.~(\ref{eq:profile_hut_epszz}) is a contribution
of the 9th edge, and the arctangent term in
Eq.~(\ref{eq:profile_hut_epsxx}) comes from the first and the third
faces. All the rest terms are similar to that of
Eq.~(\ref{eq:profile_pyramid}).

\begin{figure}
\includegraphics[width=3in]{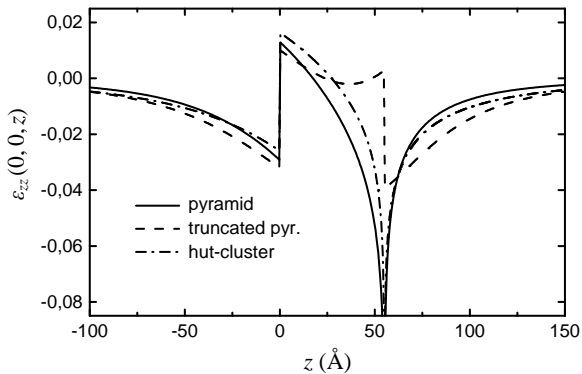}
\caption{\label{fig:profile} The strain component,
$\varepsilon_{zz}$, plotted along the $z$ axis for a pyramid
(Eq.~(\ref{eq:profile_pyramid}), solid line), a truncated pyramid
(Eq.~(\ref{eq:profile_truncated}), dash line), and a hut-cluster
(Eq.~(\ref{eq:profile_hut_epszz}), dash-dot line).}
\end{figure}

As an illustration, in Fig.~\ref{fig:profile} we plotted profiles of
the strain component, $\varepsilon_{zz}$, calculated by
Equations~(\ref{eq:profile_pyramid}--\ref{eq:profile_hut_epszz}) for
some particular cases of pyramidal, truncated pyramidal, and
hut-cluster inclusions. Parameters of the structures are chosen to
be the same as ones in Ref.~\onlinecite{Pearson2000_pyramid}:
$\varepsilon_0=0.067$; $\nu=0.24$; $a=155$~\AA; $h=55$~\AA\ for the
pyramid and the hut-cluster; $h=110$~\AA\ and $\lambda=0.5$ for the
truncated pyramid; $b=2a$ for the hut-cluster. Presented curves for
the pyramid and the truncated pyramid are identical to ones of
Ref.~\onlinecite{Pearson2000_pyramid} (curves D and B in Fig.~5,
correspondingly), that confirms the correctness of our formulas.

A detailed discussion of these profiles is beyond the scope of the
present paper. We only note that $\varepsilon_{zz}$ diverges
logarithmically at $z=55$~\AA\ in the pyramid and the hut-cluster.
This divergence is a common feature of a strain distribution in a
vicinity of a vertex or an edge of any polyhedral inclusion.

\section{Elastic anisotropy}\label{sec:anizo}

The above consideration was based on an assumption of elastically isotropic inclusion and matrix. For applications to semiconductor heterostructures, this assumption may be a source of considerable error. For example, the Young modulus of silicon in $\langle111\rangle$ direction is 1.44 times greater than that in $\langle100\rangle$ direction. Therefore, taking the elastic anisotropy into account is an actual problem.

In this Section, we argue that our method can be expanded to anisotropic media. We start from expression of strain tensor via Green's tensor $G_{\alpha\beta}$ by Faux and Pearson~\cite{Faux2000_expansion}:
\[
\varepsilon_{\alpha\beta}(\mathbf{r})=\varepsilon_0\int_V G_{\alpha\beta}(\mathbf{r}-\mathbf{r}')d\mathbf{r}',
\]
where $V$ is the inclusion volume, and $\varepsilon_0$ is lattice mismatch. These authors found a series expansion for Green's tensor, assuming cubic anisotropy:
\begin{equation}\label{eq:expansion}
G_{\alpha\beta} = G_{\alpha\beta}^{(0)} + \Delta G_{\alpha\beta}^{(1)} + \Delta^2 G_{\alpha\beta}^{(2)} + ...\, ,
\end{equation}
where expansion coefficient $\Delta=(C_{11}-C_{12}-2C_{44})/(C_{12}+2C_{44})$ is a measure of anisotropy ($\Delta\approx-\frac13$ for typical semiconductors), $C_{11}$, $C_{12}$ and $C_{44}$ are elastic moduli. Each term of this expansion can be presented as a combination of partial derivatives of expressions like $1/r$, $x^2/r$, etc. For example, isotropic term is
\[
G_{\alpha\beta}^{(0)}(\mathbf{r}) = -\frac{\varepsilon_0}{4\pi}\, \frac{3C_{12}+2C_{44}}{C_{12}+2C_{44}}\,
\frac{\partial^2}{\partial x_\alpha \partial x_\beta}\, \frac 1r \, ;
\]
$xx$-component of first-order correction, $G_{xx}^{(1)}$, is a linear combination of the following terms:
\[
\frac{\partial^2}{\partial x^2}\,\frac 1r, \quad
\frac{\partial^3}{\partial x^3}\,\frac xr, \quad
\frac{\partial^4}{\partial x^4}\,\frac {x^2}r, \quad
\frac{\partial^4}{\partial x^2 \partial y^2}\,\frac {y^2}r, \quad
\frac{\partial^4}{\partial x^2 \partial z^2}\,\frac {z^2}r.
\]
As a result, strain tensor $\varepsilon_{\alpha\beta}({\mathbf r})$ expresses as a combination of derivatives
\begin{equation}\label{eq:integrals}
\frac{\partial^{\,a+b+c}}{\partial x^a\,\partial y^b\,\partial z^c} \int_V (x')^d (y')^e (z')^f \frac{d\mathbf{r}'}{\mathbf{r}-\mathbf{r}'}\, ,
\end{equation}
where $a,b,c,d,e,f=0,1,2,...$ with a constraint $a+b+c=d+e+f+2$. Each term in the expansion (\ref{eq:expansion}) is a sum of a \emph{finite} number of derivatives (\ref{eq:integrals}) taken with proper constant coefficients.

According to our method, the integrals in Eq.~(\ref{eq:integrals}) can be regarded as electrostatic potentials induced by a \emph{non-uniformly} charged inclusion. Taking the derivatives in Eq.~(\ref{eq:integrals}), one proceeds from ``volume charge'' to ``surface dipoles'' on faces of the inclusion surface, and to ``linear charges'' and ``multipoles'' on its edges. Thus, our method allows to split the strain tensor into contributions of faces and edges of inclusion surface (assuming that inclusion shape is a polyhedron):
\begin{equation}\label{eq:splitting}
\varepsilon_{\alpha\beta}(\mathbf{r}) = \sum_i A_{\alpha\beta}^{(i)}(\mathbf{r}) + \sum_k B_{\alpha\beta}^{(k)}(\mathbf{r}),
\end{equation}
where indices $i$ and $k$ run over all faces and edges, correspondingly.

Explicit formulas for contributions $A_{\alpha\beta}^{(i)}$ and $B_{\alpha\beta}^{(k)}$ are beyond the scope of the present paper and are the subject of a separate publication. There we only note that each face contribution $A_{\alpha\beta}^{(i)}(\mathbf{r})$ is proportional to a solid angle $\Omega_i(\mathbf{r})$ with a coefficient depending on the orientation of this face and on elastic constants. Edge contributions $B_{\alpha\beta}^{(k)}(\mathbf{r})$ can be expressed in a closed form for each term of series expansion (\ref{eq:expansion}).

\section{Conclusions}\label{sec:concl}

In summary, we propose a new, more simple and flexible
expression for strain field in and around an inclusion buried in an
infinite or semi-infinite isotropic medium. This expression was also implemented as a computer program.\cite{easystrain} We show that the strain
field can be presented as a sum of contributions of the faces and
edges. This is the main point of our method; it gives a possibility
to construct expressions for strain distribution in inclusions of
complicated shapes. The general solution is applied to important
particular cases of pyramidal and hut-cluster inclusions. Our
solution for the pyramid reproduces previous solutions, but in a
simpler and intuitively understandable form. We believe that it
paves the way for further simplifications and generalizations of the
solution, for example, to the case of anisotropic elasticity.

\begin{acknowledgments}
This work was supported by RFBR (grant 06-02-16988), the Dynasty
foundation, and the President's program for young scientists (grant
MK-4655.2006.2).
\end{acknowledgments}

\appendix
\section{Evaluation of solid angles}\label{sec:solid}

Here we present some explicit formulas expressing solid angles as
functions of coordinates.

A solid angle $\Omega(\mathbf r)$, that a surface $S$ subtends at a
point $\mathbf r$, may be defined as an integral over the surface:
\begin{equation}\label{eq:solid_def}
\Omega(\mathbf r)=\int\limits_S\frac{\mathbf n(\mathbf r-\mathbf
r')} {|\mathbf r-\mathbf r'|^3}\,dS,
\end{equation}
where $dS$ is a surface element, $\mathbf n$ is a unit vector
directed normally to this surface element, and $\mathbf r'$ is a
position vector of the surface element.

\begin{figure}
\includegraphics[width=2.3in]{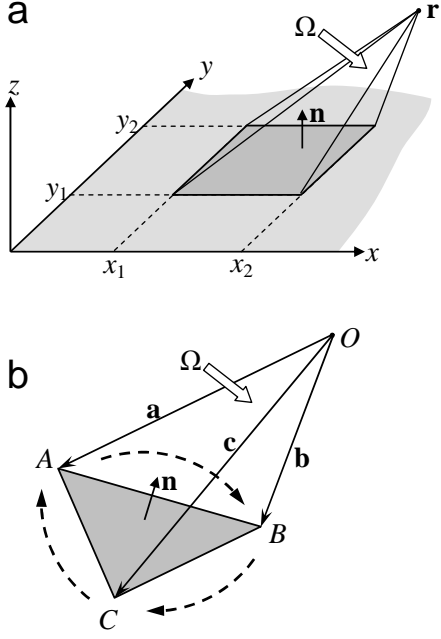}
\caption{\label{fig:solid} Solid angles: (a) expressed by
Eq.~(\ref{eq:solid_rect}); (b) expressed by
Eq.~(\ref{eq:solid_triangle}).}
\end{figure}

This integral is easily evaluated if the surface $S$ is a rectangle.
For simplicity, let this rectangle lie in the plane $z=0$, and its
edges be oriented along the axes $x$ and $y$
(Fig.~\ref{fig:solid}a). Let $x_1$ and $x_2$ be $x$-coordinates of
edges directed along the axis $y$ $(x_1<x_2)$; $y_1$ and $y_2$ be
$y$-coordinates of the rest two edges of the rectangle $(y_1<y_2)$.
Then the integral (\ref{eq:solid_def}) is expressed as follows:
\[
\Omega^\mathrm{rect}(\mathbf{r};x_1,x_2,y_1,y_2)=
\int\limits_{x_1}^{x_2}\!\!dx'\!\!\! \int\limits_{y_1}^{y_2}\!\!\!
\frac{dy'\,z}{((x\!-\!x')^2\!+\!(y\!-\!y')^2\!+\!z^2)^{3/2}},
\]
or
\begin{eqnarray}\label{eq:solid_rect}
\nonumber \lefteqn{\Omega^\mathrm{rect}(\mathbf{r};x_1,x_2,y_1,y_2)=}\\
\nonumber &&\textstyle\arctan\frac{(x-x_1)(y-y_1)}{z\,r_{11}} -
\arctan\frac{(x-x_1)(y-y_2)}{z\,r_{12}} -\\
&&\textstyle\arctan\frac{(x-x_2)(y-y_1)}{z\,r_{21}} +
\arctan\frac{(x-x_2)(y-y_2)}{z\,r_{22}}.
\end{eqnarray}
Here $r_{11}...r_{22}$ are distances from the point $\mathbf r$ to
the corners of the rectangle:
\begin{eqnarray*}
\scriptstyle r_{11}=\sqrt{(x-x_1)^2+(y-y_1)^2+z^2},\quad
r_{12}=\sqrt{(x-x_1)^2+(y-y_2)^2+z^2},\\
\scriptstyle r_{21}=\sqrt{(x-x_2)^2+(y-y_1)^2+z^2},\quad
r_{22}=\sqrt{(x-x_2)^2+(y-y_2)^2+z^2}.
\end{eqnarray*}
It is assumed that values of arctangents fall into the range
$(-\frac\pi2,+\frac\pi2)$.

To find a solid angle subtended by a triangle
(Fig.~\ref{fig:solid}b), one can use the relation by Oosterom and
Strackee: \cite{Oosterom_solid,wiki_solid}
\begin{equation}\label{eq:solid_Oosterom}
\tan\!\frac{\Omega^\mathrm{triangle}(\mathbf{a},\!\mathbf{b},\!\mathbf{c})}{2}=
\frac{[\mathbf{a}\times\mathbf{b}]\,\mathbf{c}}
{abc\!+\!(\mathbf{a}\mathbf{b})c\!+\!(\mathbf{a}\mathbf{c})b\!+\!(\mathbf{b}\mathbf{c})a}.
\end{equation}

Vectors $\mathbf a$, $\mathbf b$, $\mathbf c$ join the point $O$, at
which the solid angle is subtended, to the vertices $A$, $B$, $C$ of
the triangle. To satisfy the condition that the solid angle $\Omega$
is positive if it is looked at from outside, and is negative
otherwise, one should choose the proper order of following of the
vertices $A$, $B$, $C$. Namely, the closed contour $ABCA$ should
follow in a clockwise direction, seeing from outside the inclusion
(as shown by dashed arrows in Fig.~\ref{fig:solid}b). So the triple
scalar product $[\mathbf{a}\times\mathbf{b}]\,\mathbf{c}$ is
positive (negative) if the outer (inner) side of the triangular face
is seen from the point $O$.

Care must be taken while resolving Eq.~(\ref{eq:solid_Oosterom})
with respect to $\Omega$. For simplicity, we will refer to the right
side of Eq.~(\ref{eq:solid_Oosterom}) as to $\lambda$. The sign of
$\Omega$ is the same as the sign of the product
$[\mathbf{a}\times\mathbf{b}]\,\mathbf{c}$, but may differ from the
sign of $\lambda$. So we cannot ``naively'' resolve
Eq.~(\ref{eq:solid_Oosterom}) as $\Omega=2\arctan\lambda$. Instead,
we should write down $\Omega=2(\arctan\lambda\;\mathrm{mod}\;\pi)$
if $[\mathbf{a}\times\mathbf{b}]\,\mathbf{c}>0$, and
$\Omega=2(\arctan\lambda\;\mathrm{mod}\;\pi)-2\pi$ otherwise.
Joining together both cases, we obtain
\begin{eqnarray}\label{eq:solid_triangle}
\nonumber
\lefteqn{\Omega^\mathrm{triangle}(\mathbf{a},\!\mathbf{b},\!\mathbf{c})}\\
\nonumber
&=&2\left(\arctan\frac{[\mathbf{a}\times\mathbf{b}]\,\mathbf{c}}
{abc\!+\!(\mathbf{a}\mathbf{b}\!)c\!+\!(\mathbf{a}\mathbf{c}\!)b\!+\!(\mathbf{b}\mathbf{c}\!)a}
\,\mathrm{mod}\,\pi\right)\\
&-&2\pi\left(1-\theta([\mathbf{a}\times\mathbf{b}]\,\mathbf{c})\right),
\end{eqnarray}
where $\theta(x)$ is the Heaviside function (1 for positive $x$, 0
for negative $x$). Note that the easiest way to implement
Eq.~(\ref{eq:solid_triangle}) in a computer program is to use C math
library function atan2: $\Omega=2*\mathrm{atan2}(P,Q)$, where $P$
and $Q$ are numerator and denominator of the right part of
Eq.~(\ref{eq:solid_Oosterom}).

With Eq.~(\ref{eq:solid_triangle}), one can calculate a solid angle
subtended by an arbitrary polygon, breaking this polygon down into
triangles.

Now let us apply the expressions (\ref{eq:solid_rect}) and
(\ref{eq:solid_triangle}) to faces of the pyramid and the
hut-cluster considered in Section~\ref{sec:pyr_and_hut}. First, we
choose a reference frame with the origin at the center of the
pyramid base, the axes $x$ and $y$ along edges of the base, and the
axis $z$ directed toward the apex of the pyramid. So, position
vectors of the vertices (see Fig.~\ref{fig:pyr_hat}a) are
\begin{eqnarray*}
\mathbf r_1&\!=\!&(a/2,a/2,0),\; \mathbf r_2=(-a/2,a/2,0),\\
\mathbf r_3&\!=\!&(-a/2,-a/2,0),\; \mathbf r_4=(a/2,-a/2,0),\;
\mathbf r_5=(0,0,h),
\end{eqnarray*}
where $a$ is a base edge length, $h$ is a height of the pyramid.
Note that a dihedral angle $\vartheta=\arctan(2h/a)$. Solid angles
contributing into Eq.~(\ref{eq:pyramid_solution}) are
\begin{eqnarray}\label{eq:solid_pyramid}
\nonumber\Omega_0(\mathbf r)&=&
-\Omega^\mathrm{rect}\textstyle(\mathbf{r};-\frac a2,\frac a2,-\frac a2,\frac a2),\\
\nonumber\Omega_1(\mathbf r)&=&
\Omega^\mathrm{triangle}(\mathbf{r}_1-\mathbf{r},\mathbf{r}_4-\mathbf{r},\mathbf{r}_5-\mathbf{r}),\\
\Omega_2(\mathbf r)&=&
\Omega^\mathrm{triangle}(\mathbf{r}_2-\mathbf{r},\mathbf{r}_1-\mathbf{r},\mathbf{r}_5-\mathbf{r}),\\
\nonumber\Omega_3(\mathbf r)&=&
\Omega^\mathrm{triangle}(\mathbf{r}_3-\mathbf{r},\mathbf{r}_2-\mathbf{r},\mathbf{r}_5-\mathbf{r}),\\
\nonumber\Omega_4(\mathbf r)&=&
\Omega^\mathrm{triangle}(\mathbf{r}_4-\mathbf{r},\mathbf{r}_3-\mathbf{r},\mathbf{r}_5-\mathbf{r}).
\end{eqnarray}
The minus sign at $\Omega^\mathrm{rect}$ in
Eq.~(\ref{eq:solid_pyramid}) reflects the fact that the pyramid base
is directed downwards.

For the hut-cluster, position vectors of vertices are
\begin{eqnarray*}
\mathbf r_1&\!=\!&(b/2,a/2,0),\; \mathbf r_2=(-b/2,a/2,0),\\
\mathbf r_3&\!=\!&(-b/2,-a/2,0),\; \mathbf r_4=(b/2,-a/2,0),\\
\mathbf r_5&\!=\!&((b-a)/2,0,h),\; \mathbf r_6=(-(b-a)/2,0,h).
\end{eqnarray*}
Here $a$ and $b$ are the smaller and the bigger edge lengths of the
base. (Again, the dihedral angle is $\vartheta=\arctan(2h/a)$.) We
introduce also two points $O_1$ and $O_2$ where side edges cross
(see Fig.~\ref{fig:pyr_hat}b):
\[
\mathbf r_{O_1}=(0,-(b-a)/2,hb/a),\; \mathbf
r_{O_2}=(0,(b-a)/2,hb/a).
\]
The solid angle $\Omega_2$ of the trapezoidal 2nd face is the
difference of two solid angles subtended by triangles $O_112$ and
$O_156$. The angle $\Omega_4$ is evaluated in the same way. As a
result, solid angles subtended by faces of the hut-cluster are
\begin{eqnarray}\label{eq:solid_hut}
\nonumber\Omega_0(\mathbf r)&=&
-\Omega^\mathrm{rect}\textstyle(\mathbf{r};-\frac b2,\frac b2,-\frac a2,\frac a2),\\
\nonumber\Omega_1(\mathbf r)&=&
\Omega^\mathrm{triangle}(\mathbf{r}_1-\mathbf{r},\mathbf{r}_4-\mathbf{r},\mathbf{r}_5-\mathbf{r}),\\
\nonumber\Omega_2(\mathbf r)&=&
\Omega^\mathrm{triangle}(\mathbf{r}_2-\mathbf{r},\mathbf{r}_1-\mathbf{r},\mathbf{r}_{O_1}-\mathbf{r})\\
&-&\Omega^\mathrm{triangle}(\mathbf{r}_6-\mathbf{r},\mathbf{r}_5-\mathbf{r},\mathbf{r}_{O_1}-\mathbf{r}),\\
\nonumber\Omega_3(\mathbf r)&=&
\Omega^\mathrm{triangle}(\mathbf{r}_3-\mathbf{r},\mathbf{r}_2-\mathbf{r},\mathbf{r}_6-\mathbf{r}),\\
\nonumber\Omega_4(\mathbf r)&=&
\Omega^\mathrm{triangle}(\mathbf{r}_4-\mathbf{r},\mathbf{r}_3-\mathbf{r},\mathbf{r}_{O_2}-\mathbf{r})\\
\nonumber&-&\Omega^\mathrm{triangle}(\mathbf{r}_5-\mathbf{r},\mathbf{r}_6-\mathbf{r},\mathbf{r}_{O_2}-\mathbf{r}).
\end{eqnarray}
These expressions can be substituted into a modified version of
Eq.~(\ref{eq:pyramid_solution}), as described in
Section~\ref{sec:pyr_and_hut}.


\begin{thebibliography}{00}

\bibitem{Bir_Pikus}
G. L. Bir and G. E. Pikus, \emph{Symmetry and Strain-Induced Effects
in Semiconductors} (Wiley, New York, 1974).

\bibitem{Van_de_Walle}
C. G. Van de Walle, Phys. Rev. B \textbf{39}, 1871 (1989).

\bibitem{Yakimov}
A. V. Dvurechenskii, A. V. Nenashev, and A. I. Yakimov,
Nanotechnology \textbf{13}, 75 (2002).

\bibitem{Stangl2004_RMP}
J. Stangl, V. Hol\'y, and G. Bauer, Rev. Mod. Phys. \textbf{76}, 725
(2004).

\bibitem{Maranganti2006_TCN}
R. Maranganti and P. Sharma, \emph{Handbook of Theoretical and
Computational Nanotechnology}, Chapter 118 (2006).

\bibitem{Grundmann1995}
M. Grundmann, O. Stier, and D. Bimberg, Phys. Rev. B \textbf{52},
11969 (1995).

\bibitem{Pryor1998}
C. Pryor, Phys. Rev. B \textbf{57}, 7190 (1998).

\bibitem{Stier1999}
O. Stier, M. Grundmann, and D. Bimberg, Phys. Rev. B \textbf{59},
5688 (1999).

\bibitem{Christiansen1994}
S. Christiansen, M. Albrecht, H. P. Strunk, and H. J. Maier, Appl.
Phys. Lett. \textbf{64}, 3617 (1994).

\bibitem{Noda1998}
S. Noda, T. Abe, and M. Tamura, Phys. Rev. B \textbf{58} 7181
(1998).

\bibitem{Cusack1996}
M. A. Cusack, P. R. Briddon, and M. Jaros, Phys. Rev. B \textbf{54},
R2300 (1996).

\bibitem{Nenashev2000}
A. V. Nenashev and A. V. Dvurechenskii, Zh. Eksp. Teor. Fiz.
\textbf{118}, 570 (2000) [Engl. transl. JETP \textbf{91}, 497
(2000)].

\bibitem{Kikuchi2001}
Y. Kikuchi, H. Sugii, and K. Shintani, J. Appl. Phys. \textbf{89},
1191 (2001).

\bibitem{Daruka1999}
I.~Daruka, A.-L.~Barabasi, S.~J.~Zhou, T.~C.~Germann, P.~S.~Lomdahl,
and A.~R.~Bishop, Phys. Rev. B \textbf{60}, R2150 (1999).

\bibitem{Faux1996}
D. A. Faux, J. R. Downes, and E. P. O'Reilly, J. Appl. Phys.
\textbf{80}, 2515 (1996).

\bibitem{Downes1997_cuboid}
J. R. Downes, D. A. Faux, and E. P. O'Reilly, J. Appl. Phys.
\textbf{81}, 6700 (1997).

\bibitem{Stoleru2002_pyramid}
V. G. Stoleru, D. Pal, and E. Towe, Physica E \textbf{15}, 131
(2002).

\bibitem{Andreev1999}
A. D. Andreev, J. R. Downes, D. A. Faux, and E. P. O'Reilly, J.
Appl. Phys. \textbf{86}, 297 (1999).

\bibitem{Glas2001}
F. Glas, J. Appl. Phys. \textbf{90}, 3232 (2001).

\bibitem{Eshelby1957}
J. D. Eshelby, Proc. R. Soc. London, Ser. A \textbf{241}, 376
(1957).

\bibitem{Davies1998}
J. H. Davies, J. Appl. Phys. \textbf{84}, 1358 (1998).

\bibitem{Pearson2000_pyramid}
G. S. Pearson and D. A. Faux, J. Appl. Phys. \textbf{88}, 730
(2000).

\bibitem{Faux1997_wire}
D. A. Faux, J. R. Downes, and E. P. O'Reilly, J. Appl. Phys.
\textbf{82}, 3754 (1997).

\bibitem{Nozaki2001_general_solution}
H. Nozaki and M. Taya, J. Appl. Mech. \textbf{68}, 441 (2001).

\bibitem{Davies2003_semi_inf}
J. H. Davies, J. Appl. Mech. \textbf{70}, 655 (2003).

\bibitem{Tamm}
I. E. Tamm, \emph{Fundamentals of the Theory of Electricity} (Mir,
Moscow, 1979), p. 82.

\bibitem{Landau}
L. D. Landau and E. M. Lifshitz, \emph{Theory of Elasticity}.

\bibitem{Faux2000_expansion}
D.A. Faux and G.S. Pearson, Phys. Rev. B \textbf{62}, 4798 (2000).

\bibitem{easystrain}
http://easystrain.narod.ru

\bibitem{Oosterom_solid}
A. Van Oosterom and J. Strackee, IEEE Trans. on Biomed. Eng.
\textbf{30}, 125 (1983).

\bibitem{wiki_solid}
http://en.wikipedia.org/wiki/Solid\_angle

\end{thebibliography}
\end{document}